\documentclass[12pt,preprint]{aastex}
\newcommand{\beq}{\begin{equation}}
\newcommand{\beqa}{\begin{eqnarray}}
\newcommand{\eeq}{\end{equation}}
\newcommand{\eeqa}{\end{eqnarray}}

\def\sj{{\scriptscriptstyle (j)}}
\def\tot{{{\rm tot}}}
\def\sub{{{\rm sub}}}
\def\obs{{{\rm obs}}}
\def\dep{{{\rm dep}}}
\def\dur{{{\rm dur}}}

\slugcomment{published in ApJL}
\shorttitle{Prompt Optical Emission of GRBs}
\shortauthors{Doi, Takami, \& Yamazaki}
\begin{document}

\title{A Unified Model of the Prompt Optical Emission
of Gamma-Ray Bursts}
\author{Hirotsugu Doi\altaffilmark{1},
Kentaro Takami\altaffilmark{1}, 
and
Ryo Yamazaki\altaffilmark{1}
}
%% \email{ryo@theo.phys.sci.hiroshima-u.ac.jp}

\altaffiltext{1}{
Department of Physical Science, Hiroshima University, Higashi-Hiroshima,
Hiroshima 739-8526, Japan;
doi@theo.phys.sci.hiroshima-u.ac.jp,
takami@theo.phys.sci.hiroshima-u.ac.jp,
ryo@theo.phys.sci.hiroshima-u.ac.jp.
}

%\date{April 18, 2001}

\begin{abstract}
The observational diversity of optical emission,
which coincides with prompt
 gamma-ray bursts (GRBs), has been discovered
in the recent {\it Swift} era.
We show that on the assumption of the synchrotron radiation
for the observed energy  range below the X-ray band,
the observed diversity
 can be explained using the internal shock model by 
taking into account
a high-latitude emission and the spectral change due to the
synchrotron self-absorption.
It may even be possible in our model to include bright optical flashes 
found, e.g., in GRB~990123.
The prediction of our model is that 
the spectral index in the optical band 
is dependent on whether  the optical light curve correlates with  those
in the X-rays and/or $\gamma$-rays or not,
which will be tested in the near future.

\end{abstract}

\keywords{gamma rays: bursts --- gamma rays: theory}
%\pacs{PACS numbers:  }
%\vfill\eject

%\baselineskip 8mm

\section{Introduction}
\label{sec:intro}

Thanks to  {\it Swift} observations,
$\gamma$-ray bursts (GRBs) can be
rapidly followed-up in various observation bands
\citep[see][for a recent review]{zhang07}.
In particular, some events have been observed in the
optical band during the prompt $\gamma$-ray--active
phase \citep[e.g.,][]{yost2006,rykoff2006,roming06}.
There is observational diversity in the 
prompt optical emission and several categories can be found
as described in the following:

(i)~The optical light curve synchronizes with
the X-ray/$\gamma$-ray light curve with
the optical--to--X/$\gamma$-ray flux ratio almost constant with time
\citep[e.g., GRB~041219a;][]{vestrand2005}.
The peak flux ratio 
$R_{\rm peak}=F_{\nu_R}^{\rm peak}/F_{\nu_X}^{\rm peak}$
is from a few to ten.

(ii)~The optical light curve is smooth and much less
variable compared with X-ray/$\gamma$-ray light curve
\citep[e.g., GRB~060210;][]{stanek2006}.
The peak flux ratio $R_{\rm peak}$ is almost equal to or
 smaller than unity for GRB~060210.

(iii)~The optical light curve is a superposition
of a smooth component and a variable one synchronizing
with X-ray/$\gamma$-ray light curves
\citep[e.g., GRB~050820A;][]{vestrand2006} .
This may be an intermediate case between  cases
(i) and (ii).

(iv)~A few events show bright optical flashes just
after the $\gamma$-ray--active phase.
A typical example is GRB~990123 \citep{akerlof99},
and for this event
$R_{\rm peak}\sim2\times10^2$ \citep{briggs99}.
GRB~060111B may also be  in this class \citep{klotz2006}.

(v)~For a fraction of the events, UVOT on {\it Swift} detected
no optical counterpart within $10^3$~sec after the BAT trigger,
and it gave strict upper limits on the prompt optical emission
\citep{roming06}.
Although it is difficult to discuss quantitatively,
some events may have dim prompt optical emission.

The 
origin of the prompt optical emission has been widely discussed
\citep[e.g.,][]{pana00,fan2005,wei2006,panaitescu2006}.
The popular interpretation is that of case (i), which postulates
an internal shock origin \citep{vestrand2005}, while  case (iv) 
postulates an external  reverse shock emission origin
\citep[e.g.,][]{sari99,meszaros99,wang2000,zhang03,nakar2004,mcmahon2006}.
Case (iii) is a superposition of both components.
In this Letter, however, we will show that all of the cases are
explained in the internal shock model.

\section{Prompt emission model and optical/X-ray light curves}
\label{sec:model}

Basically, we adopt the same model as in the
previous works \citep{yama04b,toma05a,toma05}.
However, in this Letter, a much simplified version
is considered.
The central engine launches $N_{\rm tot}$-emitting shells
in the same direction, 
with the Lorentz factor $\gamma=(1-\beta^2)^{-1/2}$ and
the opening half-angle of $\Delta\theta_{\rm tot}$.
We introduce the spherical coordinate system
$(t,r,\vartheta,\varphi)$ in the central engine frame,
where the origin is the location of the central engine and
$\vartheta=0$ is the axis of the whole jet.
We assume that an observer sees the jet on-beam
(i.e., in notation of our previous works,
$\Delta\theta_\sub^\sj=\Delta\theta_{\rm tot}$,
$\vartheta^\sj=0$,
and
$\vartheta_\obs=\theta_v^\sj=0$, then
$\Delta\phi^\sj=\pi$).
The departure time of each emitting shell $t_{\rm dep}^\sj$
is assumed to be homogeneously random between
$t=0$ and $t=t_{\rm dur}$, where $t_{\rm dur}$ is the duration
time of the central engine in its own frame.
The jet emission instantaneously arises at $r=r_0$.
Then the
observed flux $F_\nu(T)$ at an observer time $T$ and at an
observed frequency $\nu$ is a superposition of
each shell emission calculated as \citep{toma05}
\begin{equation}
F_\nu^\sj (T)= K_0\tau_\sj{}^{-2}
f(\tau_\sj\nu/\gamma)~~,
\label{eq:flux}
\end{equation}
where
\begin{equation}
\tau_\sj=(c\gamma^2/r_0)(T-t_\dep^\sj)
\label{eq:tau}
\end{equation}
and $K_0$ is the normalization constant.
Here $T=0$ is chosen as the time of arrival at the observer
of a photon emitted at the origin at $t=0$.
In our model, synchrotron radiation is considered.
Assuming the standard internal shock model and typical model
parameters, the break frequencies at the on-beam observer
are calculated as 
$\nu_m\sim 30$~keV and 
$\nu_a\sim 10$~eV, where
 we assume the bulk Lorentz factor $\Gamma=300$, 
the Lorentz factor of the internal shock $\Gamma_{sh}=5$, 
the outflow luminosity $L_m=1\times10^{52}$~erg~s$^{-1}$, 
the observed typical variability timescale $\delta t=1$~sec,
and the shock microphysics parameters
$\varepsilon_e=\varepsilon_B=0.5$
\citep[see eqs.(1)--(3) in][]{wei2006}.
The cooling frequency at the on-beam observer
$\nu_c$ is found to be much smaller than $\nu_a$.
Note that considering several uncertainties of the parameters
arising in the internal shock model,
$\nu_a$ could be around the R-band frequency
$\nu_R=1.7$~eV.
Then the spectral shape function, $f(\nu')$, is 
given by \citep[e.g.,][]{pana00,fan2004}
\begin{equation}
f(\nu')=
\left\{
\begin{array}
{l@{} c@{}}
(\nu'/\nu'_a)^{5/2} & ~~(\nu'<\nu'_a)  \\
(\nu'/\nu'_a)^{-1/2} & ~~(\nu'_a<\nu'<\nu'_m)  \\
(\nu'_m/\nu'_a)^{-1/2}(\nu'/\nu'_m)^{-p/2} & ~~(\nu'_m<\nu')  \\
\end{array}
\right.~~,
\label{eq:spectrum}
\end{equation}
which is different from that for the case
$\nu'_a<\nu'_c<\nu'_m$ \citep{sari98}.
One should remark that for $\nu'<\nu'_a$, 
$f(\nu')$ has a steep slope 
because of the synchrotron self-absorption.
We don't consider the spectrum in the range
$\nu'<\nu'_c$ in Eq.~(\ref{eq:spectrum}), because
 we assume $\nu'_c\ll\nu'_a$ based on the above order-of-magnitude
estimation and then our result does not depend on the value of
$\nu'_c$.
In order to more accurately reproduce the observational
properties such as $R_{\rm peak}$, we may have to
introduce another emission component that is responsible for the
high-energy X-rays and/or $\gamma$-rays in the observer frame, 
whose spectrum is characterized by the Band function \citep{band93}.
However such a correction is neglected here for simplicity.
The pulse starting and ending times of each shell emission are
\begin{eqnarray}
T_{\rm start}^\sj &=& t_{\rm dep}^\sj +(r_0/\beta c)
(1-\beta)~~, \\
T_{\rm end}^\sj &=& t_{\rm dep}^\sj +
(r_0/\beta c)(1-\beta\cos\Delta\theta_{\rm tot})~~.
\end{eqnarray}
In this Letter, the cosmological effect is neglected 
(i.e., we set $z=0$).
In the following, we adopt the fiducial parameters
$r_0=1\times10^{15}$~cm, $\gamma=100$,
$\Delta\theta_\tot=0.25$~rad,
$t_{\rm dur}=10$~sec, $N_{\rm tot}=50$, 
$\nu'_m=1$~keV, and $p=2.5$ unless otherwise stated.
We will see how the result changes if we change
$\nu'_a$, which ranges between $1\times10^{-3}$~eV
and 0.1~eV.
We stress that the aim of this Letter is to
show that the observed diversity of the prompt optical emission
can  possibly be explained by our
theoretical model. Therefore, a thorough study of the 
dependence of the other parameters
 is beyond the scope of this Letter and of future work.

Before discussing further, one should note that since
$f(\nu')$ has a power-law form of $f(\nu')\propto\nu'^\beta$,
Eqs.~(\ref{eq:flux}) and (\ref{eq:tau}) provide us with
\begin{equation}
F_\nu^\sj \propto (T-t_{\rm dep}^\sj)^{-2+\beta}\nu^\beta~~,
\end{equation}
which is the same form as a well known formula that is
believed to describe the steep decay phase
in the early X-ray afterglow
\citep{kumar00,liang06,nousek06,obrien06,zhang06,yama06,butler06}.
If the absorption frequency at the observer, 
$\nu_a$ ($\nu_a\sim2\gamma\nu'_a$ if $T\sim T_{\rm start}^\sj$),
 is below $\nu_R$, then
$\beta=-1/2$ at $\nu_R$, so that
$F_\nu^\sj$ rapidly decreases with $\tau_\sj$.
On the other hand, if $\nu_a>\nu_R$, then $\beta=5/2$
at $\nu_R$, so that one can find
$F_\nu^\sj \propto \tau_\sj^{1/2}$.
Keeping this in mind, we expect that 
the optical behavior for $\nu_a<\nu_R$ is rather different 
from that for $\nu_R<\nu_a$,
and that the observed diversity of
prompt optical phenomena arises when
$\nu_a$ varies across $\nu_R$.
In this Letter, we consider $F_{\nu_X=10~{\rm keV}}$ as a representative of 
the X-ray/$\gamma$-ray light curve because as long as 
$1~{\rm keV}\lesssim\nu\lesssim100$~keV,
the light curve $F_\nu$ shows almost the same behavior.
The following five cases (i)--(v), correspond to those
in \S~\ref{sec:intro}:

(i)~Let us first consider the case $2\gamma\nu'_a<\nu_R$.
In this case, we obtain the light curves in the
R-band ($\nu_R=1.7$~eV) and the X-ray band ($\nu_X=10$~keV) as
\begin{eqnarray}
F_{\nu_R}^\sj &=& 
K_0 (\gamma\nu'_a/\nu_R)^{1/2}\tau_\sj^{-5/2} ~~,
\label{eq:lc_R1} \\
F_{\nu_X}^\sj &=&
\left\{
\begin{array}
{l@{} c@{}}
K_0 (\gamma\nu'_a/\nu_X)^{1/2}\tau_\sj^{-5/2}
 & ~~(\tau_{\rm start}<\tau_\sj<\gamma\nu'_m/\nu_X)  \\
K_0 (\nu'_a/\nu'_m)^{1/2}(\gamma\nu'_m/\nu_X)^{p/2}\tau_\sj^{-2-p/2}
 & ~~(\gamma\nu'_m/\nu_X<\tau_\sj<\tau_{\rm end} ) 
\end{array}
\right.~~,
\label{eq:lc_X}
\end{eqnarray}
where 
\begin{eqnarray}
\tau_{\rm start}&=&
(c\gamma^2/r_0)(T_{\rm start}-t_{\dep}^\sj)=1/2~~, \\
\tau_{\rm end}&=&
(c\gamma^2/r_0)(T_{\rm end}-t_{\dep}^\sj)
=[1+(\gamma\Delta\theta_{\rm tot})^2]/2 ~~,
\end{eqnarray}
and we use the approximation $1-\beta\sim1/2\gamma^2$.
Hence, both the optical and the X-ray light curves show
rapid decay after the initial sudden rise, so that
the optical behavior is similar to the X-ray behavior.
When $N_{\tot}\gg1$, we see many spikes as observed
for typical bursts. 
In particular, the time of the maximum flux is the same in the
optical and the X-ray bands.
In Figure~1a,
we chose $\nu'_a=1\times10^{-3}$~eV with other parameters
being fiducial.
The peak flux ratio is estimated as 
$R_{\rm peak}=F_{\nu_R}^{\rm peak}/F_{\nu_X}^{\rm peak}
\sim(\nu_X/\nu_R)^{1/2}\sim10^2$,
which is larger than the observed value.
If we consider the additional
Band function component which dominates 
only in the observed X-ray and $\gamma$-ray bands,
the value of $R_{\rm peak}$($\sim$ several tens) can be
closer to the observed one.

(ii)~Next we consider the case $\nu_R<2\gamma\nu'_a$.
Then we derive
\begin{eqnarray}
F_{\nu_R}^\sj = 
\left\{
\begin{array}
{l@{} c@{}}
K_0 (\nu_R/\gamma\nu'_a)^{5/2}\tau_\sj^{1/2}
 & ~~(\tau_{\rm start}<\tau_\sj<\gamma\nu'_a/\nu_R)  \\
K_0 (\gamma\nu'_a/\nu_R)^{1/2}\tau_\sj^{-5/2}
 & ~~(\gamma\nu'_a/\nu_R<\tau_\sj<\tau_{\rm end} ) 
\end{array}
\right.~~,
\label{eq:lc_R2} 
\end{eqnarray}
while the X-ray light curve is again described by Eq.~(\ref{eq:lc_X}).
The optical light curve has a gradually increasing part whose duration
is $\sim(\gamma\nu'_a/\nu_R)(r_0/c\gamma^2)$.
In particular, in the case of
$t_{\dur}\lesssim(\gamma\nu'_a/\nu_R)(r_0/c\gamma^2)$,
the difference between optical and X-ray behavior
becomes significant.
If $N_{\rm tot}\gg1$ and 
$t_{\dur}\lesssim(\gamma\nu'_a/\nu_R)(r_0/c\gamma^2)$,
all components are overlaid with each other  at the time interval
$t_{\rm dur}\lesssim T\lesssim (\gamma\nu'_a/\nu_R)(r_0/c\gamma^2)$,
resulting in a smooth light curve.
On the other hand, since an X-ray light curve has short-duration,
rapidly decaying pulses, we see less of a superposition effect than 
we do in the optical case.
In Figure~1b,
we chose $\nu'_a=0.1$~eV with other fiducial parameters.
One can see that the peak flux ratio is
$R_{\rm peak}\sim O(1)$, which is roughly consistent with the
observation.

(iii)~In the intermediate case $\nu_R\lesssim2\gamma\nu'_a$,
we can see optical pulses synchronizing with the
X-ray/$\gamma$-ray pulses overlaid by a smooth component.
In Figure~1c,
we chose $\nu'_a=0.02$~eV with other fiducial parameters.

(iv)~In the case of $\nu_R<2\gamma\nu'_a$,
our model can also reproduce bright optical flash
associated with e.g., GRB~990123.
In Figure~1d,
we chose $r_0=3\times10^{15}$~cm, $\nu'_m=1$~eV, and
$\nu'_a=0.05$~eV with other fiducial parameters.
Since $\gamma\nu'_m\ll\nu_X=10$~keV, the spectrum at
$\nu_X$ enters into the regime $\nu'_m<\nu'$
in Eq.~(\ref{eq:spectrum}), 
resulting in the smaller observed X-ray flux
$F_{\nu_X}$ compared with  cases (i)--(iii). 
Then the large value 
of observed $R_{peak}$ can be reproduced.
Our result implies that external reverse shock emission
is not necessarily required.
Recently, \citet{panaitescu2006} proposed 
a synchrotron-inverse Compton model to explain 
simultaneously the optical
flash and the $\gamma$-ray behavior of GRB~990123,
which is similar claim as ours.
Note that
since $\nu'_m=1$~eV, an additional Band function component
is necessary in order to match the observed spectrum
in the range $\nu>\nu_X$.
Such a correction can be possible without changing
the value of $R_{\rm peak}\sim10^2$.

(v)~Finally, if $\nu_R\ll2\gamma\nu'_a$, then the observed
flux in the optical band becomes dim because of the steep
slope below $\nu_a$ via the synchrotron self-absorption.

\section{Discussion}
\label{sec:discussion}

In this Letter, assuming the synchrotron radiation with
the absorption frequency $\nu_a$ around the R-band frequency $\nu_R$,
we have shown that the observed diversity of the prompt optical emission arises
when $\nu_a$ varies and that our theoretical model  explains the
observation well.
At present, the broad-band spectrum of the prompt emission of the GRB
is unknown, and several possibilities have been proposed
\citep[e.g.,][]{li2004,zou05,panaitescu2006}.
Our discussion can be generalized for such cases.
If the  comoving spectral function, $f(\nu')$,
has the following form around a break frequency 
$\nu'_b\sim\nu_R/\gamma$,
\begin{equation}
f(\nu')\thickapprox
\left\{
\begin{array}
{l@{} c@{}}
(\nu'/\nu'_b)^s & ~~(\nu'<\nu'_b)  \\
(\nu'/\nu'_b)^q & ~~(\nu'_b<\nu')  \\
\end{array}
\right.~~,
\end{equation}
where $s\gtrsim2$ and $q\lesssim0$,
then, our conclusion derived in this Letter remains
unchanged qualitatively.

Our model will be tested by observing the spectral energy index,
$\beta_O$, in the optical band.
When the optical light curve synchronizes with
 the X-ray/$\gamma$-rays (i.e., $\nu_b\thickapprox2\gamma\nu'_b<\nu_R$),
then the optical band is above the break frequency, $\nu_b$,
so that the energy index is $\beta_O\approx q\lesssim0$.
On the other hand,
if the optical light curve is rather smooth
and does not synchronize with the
X-ray/$\gamma$-rays light curves,
then the optical band is below $\nu_b$ ($\nu_R<\nu_b$).
Therefore, $\beta_O\approx s\gtrsim2$ is expected.
Therefore, multicolor observation in the optical (and/or infrared) band
is important to test our model.

\acknowledgements

We would like to thank the anonymous referee and
K.~Ioka for useful comments.
This work was supported by
 Grant-in-Aid for Scientific Research
of the Japanese Ministry of Education, Culture, Sports, Science
and Technology, 18740153 (R.~Y.).

%%%%%%%%% references %%%%%%%%%%%%%%%%%%%%%%%%%%%%%%

\begin{figure}
\plotone{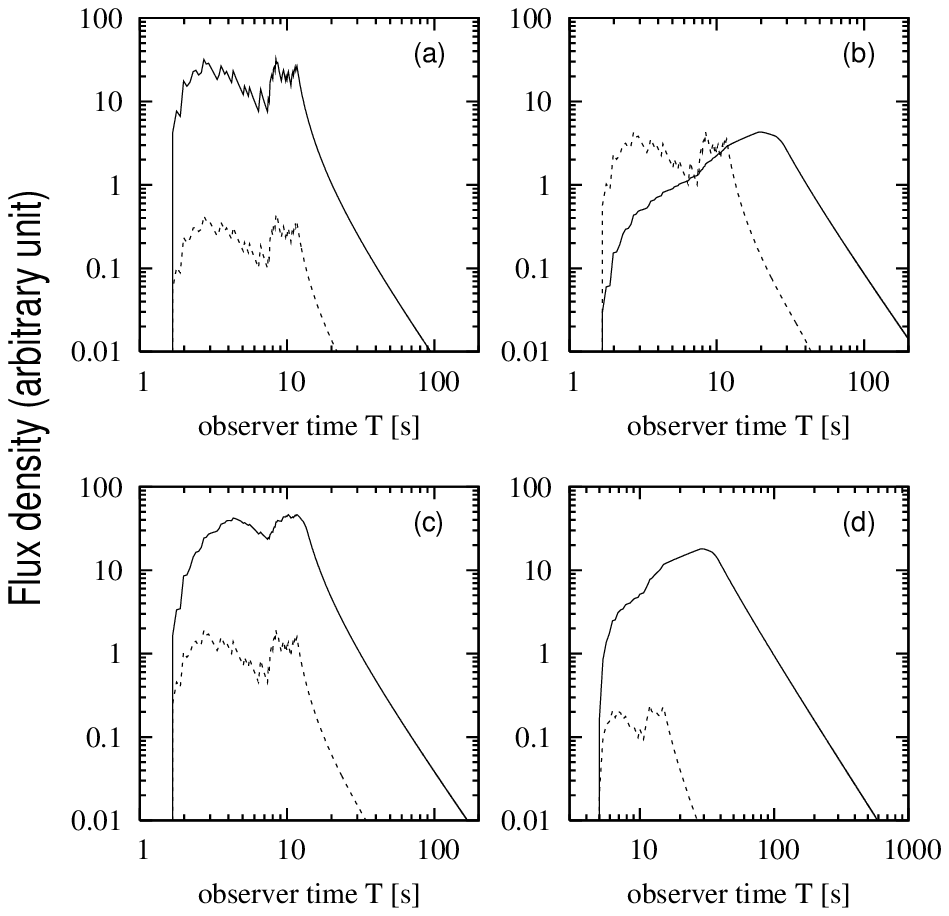}
\caption{
Examples of light curves.
Solid and dotted lines are for
 the optical ($\nu_R=1.7$~eV) and X/$\gamma$-ray 
($\nu_X=10$~keV) bands, respectively.
The adopted parameters are
(a)~$\nu'_a=1\times10^{-3}$~eV,
(b)~$\nu'_a=0.1$~eV,
(c)~$\nu'_a=0.02$~eV,
and
(d)~$\nu'_a=0.05$~eV,
$\nu'_m=1$~eV, $r_0=3\times10^{15}$~cm,
with other fiducial parameters
($r_0=1\times10^{15}$~cm, $\gamma=100$,
$\Delta\theta_\tot=0.25$~rad,
$t_{\rm dur}=10$~sec, $N_{\rm tot}=50$, 
$\nu'_m=1$~keV, and $p=2.5$).
}
\label{fig_lc}
\end{figure}


\begin{thebibliography}{99}

\bibitem[Akerlof et al.(1999)]{akerlof99}
Akerlof, C. et al. 1999, Nature, 398, 400
%
\bibitem[Band et al.(1993)]{band93}
Band,~D.~L. et al. 1993, ApJ, 413, 281
%
%\bibitem[Berger et al.(2003)]{berger03}
%Berger, E. et al., 2003, Nature, 426, 154
%
\bibitem[Briggs et al.(1999)]{briggs99}
Briggs, M. S. et al. 1999, ApJ, 524, 82
%
%% \bibitem[Burrows et al.(2005)]{burrows05}
%% Burrows, D. N. et al. 2005, Science in press, (astro-ph/0506130)
%
\bibitem[Butler \& Kocevski(2006)]{butler06}
Butler, N. R. \& Kocevski, D. 2006, astro-ph/0612564
%
%% \bibitem[Campana et al.(2005)]{campana05}
%% Campana, S. et al. 2005, ApJ, 625, L23
%
%% \bibitem[Chincarini et al.(2005)]{chincarini05}
%% Chincarini, G. et al. 2005, astro-ph/0506453
%
%% \bibitem[Connaughton(2002)]{connaughton02}
%% Connaughton, V. 2002, ApJ, 567, 1028
%
%\bibitem[Dado et al.(2006)]{dado06}
%Dado, S., Dar, A., \& De Rujula, A. 2006, ApJ, 646, L21
%
%\bibitem[Dai \& Zhang(2005)]{dai05}
%Dai, X. \& Zhang, B. 2005, ApJ, 621, 875
%
%\bibitem[Donaghy(2006)]{donaghy06}
%Donaghy, T. Q. 2006, ApJ, 645, 436
%
%\bibitem[Dyks et al.(2005)]{dyks05}
%Dyks, J., Zhang, B., \& Fan, Y. Z. 2005, astro-ph/0511699
%
%\bibitem[Eichler \& Levinson(2004)]{eichler04}
%Eichler,~D., \& Levinson,~A. 2004, ApJ, 614, L13
%
\bibitem[Fan \& Wei(2004)]{fan2004}
Fan, Y. Z. \& Wei, D. M. 2004, ApJ, 615, L69
%
\bibitem[Fan et al.(2005)]{fan2005}
Fan, Y. Z., Zhang, B., \& Wei, D. M. 2005, ApJ, 628, L25
%
%\bibitem[Ghirlanda et al.(2004)]{ghir04}
%Ghirlanda,~G., Ghisellini,~G., \& Lazzati,~D. 2004, ApJ, 616, 331
%
%% \bibitem[Giblin et al.(1999)]{giblin99}
%% Giblin, T. W. et al. 1999, ApJ, 524, L47
%
%\bibitem[Granot \& Kumar(2003)]{granot03}
%Granot,~J. \& Kumar, P. 2003, ApJ, 591, 1086
%
%% \bibitem[Granot et al.(1999)]{granot99}
%%Granot, J., Piran, T., \& Sari, R. 1999, ApJ, 513, 679
%
%% \bibitem[Ioka \& Nakamura(2001)]{ioka01}
%% Ioka,~K., \& Nakamura~T. 2001, ApJ, 554, L163
%
\bibitem[Klotz et al.(2006)]{klotz2006}
Klotz, A. et al. 2006, A\&A, 451, L39
%
%\bibitem[Kobayashi et al.(2005)]{kobayashi05}
%Kobayashi, S. \& Zhang, B. M\'{e}sz\'{a}ros,~P., \&
%Burrows, D., 2005, astro-ph/0506157
%
%\bibitem[Kobayashi \& Zhang(2006)]{kobayashi06}
%Kobayashi, S. \& Zhang, B. 2006, astro-ph/0608132
%
\bibitem[Kumar \& Panaitescu(2000)]{kumar00}
Kumar, P. \& Panaitescu, A. 2000, ApJ, 541, L51
%
%\bibitem[Kumar \& Piran(2000)]{kumar00multi}
%Kumar,~P., \& Piran,~T. 2000, ApJ, 535, 152
%
%\bibitem[Lamb et al.(2004)]{lamb04}
%Lamb,~D.~Q., et al. 2004, NewA Rev., 48, 423
%
%\bibitem[Lamb et al.(2005)]{lamb05}
%Lamb,~D.~Q., Donaghy, T. Q., \& Graziani, C. 2005, ApJ, 620, 355
%
%\bibitem[Lazzati \& Begelman(2005)]{lazzati05}
%Lazzati,~D. \& Begelman,~M.~C. 2005, ApJ, 629, 903
%
%\bibitem[Lazzati \& Begelman(2006)]{lazzati06}
%Lazzati,~D. \& Begelman,~M.~C. 2006, ApJ, 641, 972;
%
\bibitem[Li \& Song(2004)]{li2004}
Li, Z. \& Song, L. M. 2004, ApJ, 608, L17
%
\bibitem[Liang et al.(2006)]{liang06}
Liang, E. W., et al. 2006, ApJ, 646, 351
%
\bibitem[McMahon et al.(2006)]{mcmahon2006}
McMahon, E., Kumar, P., \& Piran, T. 2006, MNRAS, 366, 575
%
\bibitem[M\'{e}sz\'{a}ros \& Rees(1999)]{meszaros99}
M\'{e}sz\'{a}ros, P. \& Rees, M. 1999, MNRAS, 306, L39
%
%\bibitem[Nakamura(2000)]{nakamura00}
%Nakamura,~T. 2000, ApJ, 534, L159
%
\bibitem[Nakar \& Piran(2004)]{nakar2004}
Nakar, E. \& Piran, T. 2004, MNRAS, 353, 647
%
%\bibitem[Nakar et al.(2004)]{nakar04}
%Nakar, E.,  Granot, J., \& Guetta, D. 2004, ApJ, 606, L37
%
\bibitem[Nousek et al.(2006)]{nousek06}
Nousek, J. A. et al. 2006, ApJ, 642, 389
%
\bibitem[O'Brien et al.(2006a)]{obrien06}
O'Brien, P. T. et al. 2006a, ApJ, 647, 1213
%
%\bibitem[O'Brien et al.(2006b)]{obrien06b}
%O'Brien, P. T. et al. 2006b, astro-ph/0605230
%
\bibitem[Panaitescu \& M\'{e}sz\'{a}ros(2000)]{pana00}
Panaitescu,~A. \& M\'{e}sz\'{a}ros,~P. 2000, ApJ, 544, L17
%
%\bibitem[Panaitescu et al.(2006)]{pana06}
%Panaitescu,~A., M\'{e}sz\'{a}ros,~P., Gehrels,~N., 
%Burrows,~D., \& Nousek,~J. 2006, MNRAS, 366, 1357
%
\bibitem[Panaitescu \& Kumar(2006)]{panaitescu2006}
Panaitescu, A. \& Kumar, P. 2006, astro-ph/0612504
%
%\bibitem[Pe'er et al.(2006)]{peer06}
%Pe'er, A., M\'{e}sz\'{a}ros,~P., Rees, M. J., 2006, astro-ph/0603343
%
%\bibitem[Perna et al.(2003)]{perna03}
%Perna, R., Sari R., \& Frail, D. 2003, ApJ, 594, 379
%
\bibitem[Roming et al.(2006)]{roming06}
Roming, P. W. A. et al. 2006, ApJ, 652, 1416
%
%\bibitem[Rossi et al.(2002)]{rossi02}
%Rossi,~E., Lazzati,~D., \& Rees,~M.~J. 2002, MNRAS, 332, 945
%
\bibitem[Rykoff et al.(2006)]{rykoff2006}
Rykoff, E. S. et al. 2006, ApJ, 638, L5
%
%\bibitem[Sakamoto et al.(2005)]{sakamoto05}
%Sakamoto, T. et al. 2005, ApJ, 629, 311
%
\bibitem[Sari et al.(1998)]{sari98}
Sari, R., Piran, T., \& Narayan, R. 1998, ApJ, 497, L17
%
\bibitem[Sari \& Piran(1999)]{sari99}
Sari, R. \& Piran, T. 1999, ApJ, 520, 641
%
\bibitem[Stanek et al.(2007)]{stanek2006}
Stanek, K. Z. et al. 2007, ApJ, 654, L21
%
%\bibitem[Tagliaferri et al.(2005)]{tagliaferri06}
%Tagliaferri, G. et al. 2005, Nature, 436, 985
%
\bibitem[Toma et al.(2005a)]{toma05a}
Toma,~K., Yamazaki,~R., \& Nakamura,~T. 2005a, ApJ, 620, 835
%
\bibitem[Toma et al.(2005b)]{toma05}
Toma,~K., Yamazaki,~R., \& Nakamura,~T. 2005b, ApJ, 635, 481
%
\bibitem[Vestrand et al.(2005)]{vestrand2005}
Vestrand, W. T. et al. 2005, Nature, 435, 178
%
\bibitem[Vestrand et al.(2006)]{vestrand2006}
Vestrand, W. T. et al. 2006, Nature, 442, 172
%
\bibitem[Wang et al.(2000)]{wang2000}
Wang, W. Y., Dai, Z. G., \& Lu, T. 2000, MNRAS, 319, 1159
%
\bibitem[Wei(2007)]{wei2006}
Wei, D. M. 2007, MNRAS, 374, 525
%
%% \bibitem[Woods \& Loeb(1999)]{woods99}
%% Woods,~E. \& Loeb,~A. 1999, ApJ, 523, 187
%
%% \bibitem[Yamazaki et al.(2002)]{yama02}
%% Yamazaki,~R., Ioka,~K., \& Nakamura,~T. 2002, ApJ, 571, L31
%
%% \bibitem[Yamazaki et al.(2003)]{yama03}
%% Yamazaki,~R., Ioka,~K., \& Nakamura,~T. 2003, ApJ, 593, 941
%
%\bibitem[Yamazaki et al.(2004a)]{yama04a}
%Yamazaki,~R., Ioka,~K., \& Nakamura,~T. 2004a, ApJ, 606, L33
%
\bibitem[Yamazaki et al.(2004)]{yama04b}
Yamazaki,~R., Ioka,~K., \& Nakamura,~T. 2004, ApJ, 607, L103
%
%% \bibitem[Yamazaki et al.(2005)]{yama05}
%% Yamazaki,~R., Ioka,~K., Takahara,~F., Shibazaki,~N. 2005, PASJ, 57, L11
%
\bibitem[Yamazaki et al.(2006)]{yama06}
Yamazaki,~R., Toma, K., Ioka,~K., \& Nakamura,~T. 2006, MNRAS, 369, 311
%
%\bibitem[Yonetoku et al.(2004)]{yone04}
%Yonetoku,~D., Murakami,~T., Nakamura,~T., Yamazaki,~R., Inoue,~A.~K., \&
%Ioka,~K. 2004, ApJ, 609, 935
%
%\bibitem[Yonetoku et al.(2005)]{yone05}
%Yonetoku,~D., Yamazaki,~R., Nakamura,~T., \& Murakami,~T. 2005, MNRAS, 362, 1114
%
\bibitem[Yost et al.(2006)]{yost2006}
Yost, S. A. et al. 2006, astro-ph/0611414
%
%\bibitem[Zhang \& M\'{e}sz\'{a}ros(2002a)]{zhang02st}
%Zhang,~B., \& M\'{e}sz\'{a}ros,~P. 2002a, ApJ, 571, 876
%
%\bibitem[Zhang \& M\'{e}sz\'{a}ros(2002b)]{zhang02}
%Zhang,~B., \& M\'{e}sz\'{a}ros,~P. 2002b, Int. J. Mod. Phys. A, 19, 2385
%
\bibitem[Zhang et al.(2003)]{zhang03}
Zhang,~B., Kobayashi, S., \& M\'{e}sz\'{a}ros,~P. 2003, ApJ, 595, 950
%
%\bibitem[Zhang et al.(2004)]{zhang04}
%Zhang,~B., Dai, X., Lloyd-Ronning, N. M., \& M\'{e}sz\'{a}ros,~P. 2004, ApJ, 601, L119
%
\bibitem[Zhang et al.(2006)]{zhang06}
Zhang, B. et al. 2006, ApJ, 642, 354
%
\bibitem[Zhang(2007)]{zhang07}
Zhang, B. 2007, astro-ph/0701520
%
%\bibitem[Zhang et al.(2003)]{zhang03c}
%Zhang, W., Woosley, S. E., \& MacFadyen, A. I. 2003, ApJ, 586, 356
%
%\bibitem[Zhang et al.(2004)]{zhang04c}
%Zhang, W., Woosley, S. E., \& Heger, A. 2004, ApJ, 608, 365
%
\bibitem[Zou et al.(2006)]{zou05}
Zou, Y. C., Xu, D., \& Dai, Z. G. 2006, ApJ, 646, 1098
%
\end{thebibliography}
\end{document}